\documentclass[preprint, proceedings]{rmaa}

\usepackage{paralist}

\SetYear{2005}
\SetConfTitle{The Ninth Texas-Mexico Conference on Astrophysics}

\title{An Infrared Survey of Neutron-Capture Elements in Planetary Nebulae} 

\author{
  N. C. Sterling\altaffilmark{1}, 
  and H. L. Dinerstein\altaffilmark{1}}
\altaffiltext{1}{The University of Texas at Austin.}

\shortauthor{Sterling \& Dinerstein}
\shorttitle{Neutron-Capture Elements in Planetary Nebulae}

\fulladdresses{
\item Nicholas C. Sterling and Harriet L. Dinerstein: The University of Texas, Dept. of Astronomy, 1 University Station, C1400, Austin, TX, 78712-0259.
}

\listofauthors{N. C. Sterling, \& H. L. Dinerstein}
\indexauthor{Sterling, N. C.}
\indexauthor{Dinerstein, H. L.}

\abstract{We present results from an ongoing survey of the infrared emission lines [\ion{Kr}{3}] 2.199 and [\ion{Se}{4}] 2.287~$\mu$m in Galactic planetary nebulae (PNe).  Krypton and selenium may be produced by slow neutron-capture nucleosynthesis (the ``\emph{s}-process'') during the asymptotic giant branch (AGB) phase of PN progenitor stars, and brought to the surface by convective dredge-up before PN ejection.  We detect emission from Se and Kr in 65 of 114 Galactic PNe, and use the line fluxes to derive ionic abundances.  We employ ionization correction factors based on coincidences of ionization potentials to calculate total elemental abundances, and discuss photoionization models designed to test the veracity of these corrections.  The derived Se and Kr abundances of our targets range from approximately solar to enriched by a factor of 5, which indicates varying degrees of dredge-up and \emph{s}-process efficiencies in the progenitor stars.  In PNe exhibiting emission from both Se and Kr, we find that the relative abundances of these elements are generally in agreement with predictions from theoretical models of \emph{s}-process nucleosynthesis.  We examine our results for correlations between \emph{s}-process enrichments and other nebular properties, such as CNO abundances, morphology, and characteristics of the central stars.  PNe with Wolf-Rayet central stars tend to exhibit more elevated Se and Kr abundances than other nebulae.  Bipolar nebulae, believed to arise from the most massive of PN progenitors, may have lower \emph{n}-capture abundances than elliptical PNe.}

\addkeyword{infrared: ISM}
\addkeyword{nucleosynthesis, abundances}
\addkeyword{planetary nebulae: general}
\addkeyword{stars: AGB and post-AGB}

\begin{document}
\maketitle

\section{Introduction}

The neutron(\emph{n})-capture elements Se ($Z=34$) and Kr ($Z=36$) may be produced by the \emph{s}-process in low- and intermediate-mass stars during the thermally-pulsing phase of the asymptotic giant branch (AGB).  During this stage of evolution, free neutrons are released in the intershell region primarily by the reaction $^{13}$C($\alpha$,\emph{n})$^{16}$O, and to a lesser extent $^{22}$Ne($\alpha$,\emph{n})$^{25}$Mg.  These neutrons are captured by iron peak elements, which are transformed to heavier species by a series of \emph{n}-captures and $\beta$-decays (Busso et al.\ 1999).  The newly-produced \emph{s}-process isotopes may be conveyed to the stellar surface by third dredge-up (TDU), in which the convective envelope plunges beneath the (inactive) H-burning shell after a thermal pulse, thereby dredging up He burning products (e.g. $^{12}$C) as well as the \emph{s}-process-rich material.  TDU is therefore an important contributor to the Galactic chemical evolution of C and \emph{n}-capture elements, but it is not expected to occur in stars with masses lower than $\sim1.5$~M$_{\odot}$, and may vary in efficiency in stars with otherwise similar characteristics (Busso et al. 1999).  This mass cutoff implies that the majority of AGB stars will not undergo efficient TDU, since low-mass stars are favored by the initial mass function.

Carbon and \emph{n}-capture elements are expected to be enriched in planetary nebulae (PNe) if the progenitor star experienced TDU.  However, the abundances of these species in PNe are poorly known, as the C abundance is inherently difficult to measure in ionized nebulae (Kaler 1983; Rola \& Stasi\'{n}ska 1994), and \emph{n}-capture element abundances had been measured in only a few PNe before our survey.  In fact, due to the low abundances of trans-iron elements ($\lesssim10^{-9}$ that of H) and the consequent weakness of their spectral features, it was not until 1994 that \emph{n}-capture element lines were identified in the spectrum of a PN (P\'{e}quignot \& Baluteau 1994).  This led Dinerstein (2001) to realize that two previously anonymous features at 2.199 and 2.287 $\mu$m in the spectra of several PNe are in fact fine structure transitions of [\ion{Kr}{3}] and [\ion{Se}{4}].  Sterling et al.\ (2002) and Sterling \& Dinerstein (2003) detected Ge in absorption in the UV spectra of six PNe, and derived the Ge abundance in five of these objects, but this only expanded the number of PNe with measured \emph{n}-capture element abundances to seven.  It is not possible to determine the history of convective dredge-up and AGB nucleosynthesis in PN progenitor stars, or how these evolutionary traits are related to other observable characteristics of PNe, from such a small sample of objects.

We have conducted the first large-scale survey of \emph{n}-capture element abundances in Galactic PNe, utilizing the two near-infrared emission lines identified by Dinerstein (2001).  Our results show that [\ion{Kr}{3}] 2.199 and [\ion{Se}{4}] 2.287~$\mu$m are in fact detectable in a significant fraction ($\sim$60\%) of Galactic PNe.  Since Se and Kr are not expected to be depleted into dust (Cardelli et al.\ 1993; Cartledge et al.\ 2003), the gaseous abundances represent the true abundances of these elements.  Therefore, [\ion{Kr}{3}] 2.199 and [\ion{Se}{4}] 2.287~$\mu$m are valuable tracers of heavy element abundances and AGB nucleosynthesis in PN progenitor stars.

\section{Observations}

We have observed approximately 100 PNe in the $K$~band with the CoolSpec spectrometer (Lester et al.\ 2000) on the 2.7-m Harlan J.\ Smith Telescope at McDonald Observatory.  Each PN is observed with a $2\farcs{}7$ slit at a resolution of $\sim500$, which is sufficient to resolve [\ion{Kr}{3}] and [\ion{Se}{4}] from most nearby features.  We expand our sample by including $K$ band observations of about 15 objects from the literature (Geballe et al.\ 1991; Hora et al.\ 1999; Lumsden et al.\ 2001).  In our full sample of 114 PNe, we have detected Se and/or Kr in 65, for a detection rate of nearly 60\%.

The only important contaminants to the [\ion{Kr}{3}] and [\ion{Se}{4}] lines are vibrationally-excited lines of H$_2$, namely H$_2$~3-2~S(3) 2.201 and H$_2$~3-2~S(2) 2.287~$\mu$m.  In the 15-20\% of our targets which exhibit H$_2$ emission, we employ a high-resolution ($R\sim4400$) setting to separate [\ion{Kr}{3}] and H$_2$, and use this result to estimate the relative contribution of H$_2$ to the unresolvable blend at 2.287~$\mu$m.  In the rare case that the 2.20~$\mu$m feature is too weak to be seen in the high-resolution setting, we assume the maximum contribution of H$_2$ to the blends, e.g.\ fluorescent excitation (Model 14 of Black \& van~Dishoeck 1987).

\section{Abundances and Correlations}

We derive Kr$^{++}$ and Se$^{3+}$ ionic abundances (or upper limits) in all objects in our sample.  A five-level atom is used to determine the level populations of Kr$^{++}$, while a two-level system is used for Se$^{3+}$.  Transition probabilities and collision strengths are taken from Bi\'{e}mont \& Hansen (1986; 1987) and Sch\"{o}ning (1997) and Butler (2005, in preparation), respectively.  We especially note the recently calculated collision strength for [\ion{Se}{4}]~2.287~$\mu$m (7.71 at $10^4$~K) by K.\ Butler, which previously was unknown.  This allows us for the first time to derive absolute (rather than relative) Se abundances, and therefore we can compare the relative enrichments of Se and Kr in objects which exhibit emission from both species.

In order to determine the elemental Se and Kr abundances, it is necessary to correct for the presence of unseen stages of ionization.  We account for this by assuming that Se$^{3+}$/Se~$\approx$~Ar$^{++}$/Ar and Kr$^{++}$/Kr~$\approx$~S$^{++}$/S, based on the similar ionization potential ranges of these species.  It should be noted that this is an approximate method, and does not take into account the detailed atomic characteristics of each ion, such as the photoionization cross section and the rates of various recombination processes.  A more robust treatment of the atomic physics is possible with the aid of photoionization models, which can be used to derive more reliable ionization correction factors.  We have updated the atomic databases of the photoionization codes XSTAR (Kallman \& Bautista 2001) and CLOUDY (Ferland et al.\ 1998) to include Se and Kr.  We will use these codes to compute detailed models of about a dozen PNe (with various degrees of excitation), and derive ionization corrections from the results.  In this paper, however, we cite results using only the approximate corrections based on ionization potential energies.

\begin{table*}[!ht]\centering
\setlength{\tabnotewidth}{\columnwidth}
\tablecols{5}
\caption{Se and Kr Abundances Vs.\ Nebular Properties}
\begin{tabular}{lcccc}
\toprule
\multicolumn{1}{l}{Property} & \multicolumn{1}{c}{Derived} & \multicolumn{1}{c}{Number of}  & \multicolumn{1}{c}{Derived}  & \multicolumn{1}{c}{Number of} \\
\multicolumn{1}{l}{} & \multicolumn{1}{c}{$<$Se/H$>$} & \multicolumn{1}{c}{Se Detections}  & \multicolumn{1}{c}{$<$Kr/H$>$}  & \multicolumn{1}{c}{Kr Detections} \\
\midrule
$[$WC$]$ & 3.4$\pm$0.5 (-9) & 15 & 6.4$\pm$2.9 (-9) & 3\\
WELS & 2.8$\pm$0.7 (-9) & 9 & 3.4$\pm$1.4 (-9) & 3\\
Non-[WC]/WELS & 2.0$\pm$0.2 (-9) & 36 & 3.8$\pm$0.4 (-9) & 18\\
\midrule
Type I & 2.3$\pm$0.4 (-9) & 16 & 4.1$\pm$1.2 (-9) & 5\\
Non-Type I & 2.5$\pm$0.2 (-9) & 45 & 6.8$\pm$1.3 (-9) & 19\\
\midrule
Bipolar & 2.3$\pm$0.4 (-9) & 11 & 2.6$\pm$0.4 (-9) & 5\\
Elliptical & 3.3$\pm$0.4 (-9) & 21 & 4.2$\pm$0.7 (-9) & 11\\
Round & 2.1$\pm$0.6 (-9) & 3 & \nodata & 0\\
Irregular\tabnotemark{a} & 2.5$\pm$0.6 (-9) & 6 & 1.7 (-9) & 1\\
\midrule
Full Sample & 2.4$\pm$0.2 (-9) & 60 & 6.2$\pm$1.0 (-9) & 24\\
Solar\tabnotemark{b} & 2.14$\pm$0.20 (-9) & \nodata & 1.91$\pm$0.36 (-9) & \nodata\\
\bottomrule
\tabnotetext{a}{Includes Point-Symmetric nebulae}
\tabnotetext{b}{From Asplund et al.\ (2005)}
\end{tabular}
\end{table*}

The derived Se and Kr abundances span a wide range, from slightly subsolar (by about 0.1--0.3 dex) to enriched by a factor of five.  We interpret this range of abundances to be due to varying efficiencies of the \emph{s}-process and TDU in the progenitor stars, with the lower end of the range probably a metallicity effect.  We find that about 40\% of the 65 targets which display Se and/or Kr emission are enriched in these elements.  Converting this number to the fraction of all Galactic PN progenitors which experienced the \emph{s}-process and TDU will require a detailed study of the sample bias of our survey.

In those objects which display both Se and Kr emission, Kr tends to be more enriched than Se (see Table~1).  Theoretical models of \emph{s}-process nucleosynthesis (Busso et al.\ 2001; Busso 2003, private communication; Gallino 2005, private communication) predict that in the expected metallicity range of our sample ($\sim$0.3--1.0~Z$_{\odot}$), [Kr/Se]~=~0.0--0.5.  The exact value depends primarily on the size of the $^{13}$C pocket (i.e.\ the mass of the layer in which free neutrons are produced), which is a free parameter in the models, and to a lesser extent on the progenitor mass (with more massive progenitors having larger [Kr/Se]).  The observed [Kr/Se] ratios in our targets are generally in good agreement with the theoretical predictions, which indicates that the approximate ionization corrections we use in this paper are not likely to be wildly incorrect.

We have searched for correlations between the derived Se and Kr abundances and other nebular properties, such as CNO abundances, morphology, and central star type.  In Table~1, we display average Se and Kr abundances for a number of PN subclasses, including those with H-deficient, C-rich central stars (Wolf Rayet type, or [WC], and weak emission line stars, or WELS); PNe with large N and He enrichments (Peimbert Type~I; Peimbert 1978) which may indicate more massive progenitors ($\gtrsim$~2.5-4.0 M$_{\odot}$); and PNe of different morphological types.  We also show the average abundances for the full sample, compared to the solar abundances compiled by Asplund et al.\ (2005).

We find that PNe with [WC] central stars tend to be more enriched in \emph{n}-capture elements than other PNe.  This is true for both Se and Kr, and is likely caused in part by the deep mixing and heavy mass-loss that these stars experience as they lose their H-rich envelope (e.g.\ Bl\"{o}cker 2001).  The situation is less clear for PNe with WELS nuclei, as they seem to be more enriched in Se than PNe with H-rich central stars, but not Kr (however, the Kr results in this case are weakened by small-number statistics).  Also, bipolar PNe, which may arise from more massive progenitor stars, tend to be less enriched than ellipticals.  This indicates that the \emph{s}-process may be less efficient in more massive AGB stars.  Such a trend is also seen in Kr for Type~I vs.\ non-Type~I PNe, but the difference in Se abundances is not statistically significant.

Interestingly, no relation is seen between the Se and Kr abundances and the C/O ratio.  This comes as a surprise, since carbon is expected to be dredged up along with \emph{n}-capture elements by TDU.  Indeed, the abundances of \emph{n}-capture elements have been found to scale with C/O in AGB (Smith \& Lambert 1990; Abia et al.\ 2002) and post-AGB stars (Van~Winckel 2003).  The lack of a detected correlation in PNe may be due to uncertainties in the derived Se and Kr abundances, as well as in the C abundances (which are usually not known to better than a factor of two in PNe).

\section{Conclusions}

We report preliminary results from the first large-scale survey of \emph{n}-capture elements in Galactic PNe.  We have detected emission lines of Se and/or Kr in 65 of 114 PNe, from which we have derived the elemental Se and Kr abundances (or upper limits) for all targets in our sample.  The recently calculated [\ion{Se}{4}] 2.287~$\mu$m collision strength allows us to determine absolute Se abundances for the first time.  We find a range of [Se/H] and [Kr/H], from slightly subsolar to enriched by a factor of five relative to solar.  We interpret this as a variation in the efficiencies of the \emph{s}-process and convective dredge-up in the progenitor stars.  The abundances reported in this paper are rather uncertain, but will be made more robust as the ionization corrections are studied with the aid of the photoionization codes XSTAR and CLOUDY.  The large number of Se and Kr detections will allow us to examine AGB nucleosynthesis in PN progenitors, its efficiency in different subclasses of PNe, and consequently the Galactic chemical evolution of heavy elements.

\acknowledgements

This work has been supported by NSF grants AST 97-31156 and AST 04-06809.


\begin{thebibliography}

\bibitem{} Abia, C., et al.\ 2002, \apj, 579, 817

\bibitem{} Asplund, M., Grevesse, N., \& Sauval, A.\ J.\ 2005, in ASP Conf.\ Ser., "Cosmic Abundances As Records of Stellar Evolution and Nucleosynthesis", eds.\ F.\ N.\ Bash \& T.\ G.\ Barnes, in press (astro-ph/0410214)

\bibitem{} Bi\'{e}mont, E., \& Hansen, J.\ E.\ 1986, Phys.\ Scripta, 34, 116

\bibitem{} Bi\'{e}mont, E., \& Hansen, J.\ E.\ 1987, Nuclear Instruments and Methods in Phys.\ Res.\ B, 23, 274

\bibitem{} Black, J.\ H., \& van~Dishoeck, E.\ F.\ 1987, \apj, 322, 412

\bibitem{} Bl\"{o}cker, T.\ 2001, \apss, 275, 1

\bibitem{} Busso, M., Gallino, R., Lambert, D.\ L., Travaglio, C., \& Smith, V.\ V.\ 2001, \apj, 557, 802

\bibitem{} Busso, M., Gallino, R., \& Wasserburg, G.\ J.\ 1999, \araa, 37, 239

\bibitem{} Cardelli, J.\ A., Federman, S.\ R., Lambert, D.\ L., \& Theodosiou, C.\ E.\ 1993, \apj, 416, L41

\bibitem{} Cartledge, S.\ I.\ B., Meyer, D.\ M., \& Lauroesch, J.\ T.\ 2003, \apj, 597, 408

\bibitem{} Dinerstein, H.\ L.\ 2001, \apj, 550, L223
 
\bibitem{} Ferland, G.\ J.Ferland, Korista, K.\ T., Verner, D.\ A., Ferguson, J.\ W., Kingdon, J.\ B., \& Verner, E.\ M.\ 1998, \pasp, 110, 761

\adjustfinalcols

\bibitem{} Geballe, T.\ R., Burton, M.\ G., \& Isaacman, R.\ 1991, \mnras, 253, 75

\bibitem{} Hora, J.\ L., Latter, W.\ B., \& Deutsch, L.\ K.\ 1999, \apjs, 124, 195

\bibitem{} Kaler, J.\ B.\ 1983, in IAU Symp.\ 103, ``Planetary Nebulae,'' ed.\ D.\ R.\ Flower (Dordrecht: Reidel), 245

\bibitem{} Kallman, T., \& Bautista, M.\ 2001, \apjs, 133, 221

\bibitem{} Lester, D.\ F., Hill, G.\ J., Doppman, G., \& Froning, C.\ S.\ 2000, \pasp, 112, 384

\bibitem{} Lumsden, S.\ L., Puxley, P.\ J., \& Hoare, M.\ G.\ 2001, \mnras, 328, 419

\bibitem{} Peimbert, M.\ 1978, in IAU Symp.\ 76, ``Planetary Nebulae','' ed.\ Y.\ Terzian (Dordrecht: Reidel), 215

\bibitem{} P\'{e}quignot, D., \& Baluteau, J.-P.\ 1994, \aap, 283, 593

\bibitem{} Rola, C., \& Stasi\'{n}ska, G.\ 1994, \aap, 282, 199

\bibitem{} Sch\"{o}ning, T.\ 1997, \aaps, 122, 277

\bibitem{} Smith, V.\ V., \& Lambert, D.\ L.\ 1990, \apjs, 72, 387

\bibitem{} Sterling, N.\ C., Dinerstein, H.\ L., \& Bowers, C.\ W.\ 2002, \apj, 578, L55

\bibitem{} Sterling, N.\ C., \& Dinerstein, H.\ L.\ 2003, RMxA\&A Ser.\ de Conf., 18, 133

\bibitem{} Van Winckel, H.\ 2003, \araa, 41, 391
 
\end{thebibliography}
\end{document}